\documentstyle[epsf]{aipproc}

\catcode`@=11
\long\def\@makefntext#1{\parindent 0pt\hsize\columnwidth\parskip0pt\relax
\footnotesize\baselineskip12pt\def\strut{\vrule width0pt height0pt depth1.75pt\relax}%
\mbox{$\m@th^{\@thefnmark}$\hspace*{3pt}}#1}

\def\title#1{\gdef\@title{{\par\vskip-10pt\Large\bf
\baselineskip20pt\centering\ignorespaces#1\vskip6pt}}%
\setcounter{part}{0}
\setcounter{table}{0}
\setcounter{figure}{0}
\setcounter{equation}{0}
\setcounter{section}{0}
\setcounter{subsection}{0}
\setcounter{subsubsection}{0}
\setcounter{paragraph}{0}
}

\def\author#1{\expandafter\def\expandafter\@authoraddress\expandafter
{\@authoraddress %
{\dimen0=-\prevdepth \advance\dimen0 by1.2\baselineskip
\nointerlineskip \centering
\vrule height\dimen0 width0pt\relax\ignorespaces\large\rm#1\par
}%
}%
}

\def\address#1{\expandafter\def\expandafter\@authoraddress\expandafter
{\@authoraddress{\nointerlineskip\vskip1pc
                 \footnotesize\it\centering\ignorespaces#1\par}}}

\catcode`@=12

\def\E{\,\rlap/\!E_T}
\def\srts{\sigma_{\!\!\!\sqrt s}^{\vphantom y}}

\newcommand{\lsim}{\mathrel{\raisebox{-.6ex}{$\stackrel{\textstyle<}{\sim}$}}}
\newcommand{\gsim}{\mathrel{\raisebox{-.6ex}{$\stackrel{\textstyle>}{\sim}$}}}
\begin{document}

\font\fortssbx=cmssbx10 scaled \magstep2
\hbox to \hsize{
%\special{psfile=uwlogo.ps hscale=6000 vscale=6000 hoffset=-12 voffset=-2}
%\hskip.35in \raise.1in
\hbox{\fortssbx University of Wisconsin - Madison}
\hfill$\vcenter{\normalsize\hbox{\bf MADPH-98-1040}
                \hbox{March 1998}}$ }

\title{Overview of Physics at a Muon Collider\footnote{Invited talk presented at the {\it 4th International Conference on the Physics Potential and Development of $\mu^+\mu^-$ Colliders}, San Francisco, December 1997} }
\author{V. Barger}
\address{Physics Department, University of Wisconsin, Madison, WI 53706, USA}

\maketitle

\pagestyle{plain}
\thispagestyle{empty}

\vglue-4ex

\begin{abstract}
Muon colliders offer special opportunities to discover and study new physics. With the high intensity source of muons at the front end, orders of magnitude improvements would be realized in searches for rare muon processes, in deep inelastic muon and neutrino scattering experiments, and in long-baseline neutrino oscillation experiments. At a 100 to 500~GeV muon collider, neutral Higgs boson (or techni-particle) masses, widths and couplings could be precisely measured via $s$-channel production. Also, threshold cross-section studies of $W^+W^-$, $t\bar t$, $Zh$ and supersymmetric particle pairs would precisely determine the corrresponding masses and test supersymmetric radiative corrections. At the high energy frontier a 3 to 4~TeV muon collider is ideally suited for the study of scalar supersymmetric particles and extra $Z$-bosons or strong $WW$ scattering.
\end{abstract}

\section{Introduction}

The agenda of physics at a muon collider falls into three categories: front end physics with a high-intensity muon source, First Muon Collider (FMC) physics at a machine with center-of-mass energies of 100 to 500~GeV, and Next Muon Collider (NMC) physics at 3--4~TeV center-of-mass energies.

At the front end, a high-intensity muon source will permit searches for rare muon processes at branching sensitivities that are orders of magnitude below present upper limits. Also, a high-energy muon-proton collider can be constructed to probe high $Q^2$ phenomena beyond the reach of the HERA $ep$ collider. In addition, the decaying muons will provide high-intensity neutrino beams for precision neutrino cross-section measurements and for long-baseline experiments. 

The FMC will be a unique facility for neutral Higgs boson (or techni-resonance) studies through $s$-channel resonance production. Measurements can also be made of the threshold cross sections for $W^+W^-$, $t\bar t$, $Zh$, $\chi_1^+\chi_1^-$, $\chi_2^0\chi_1^0$, $\tilde\ell^+\tilde\ell^-$ and $\tilde\nu\tilde\nu$ production that will determine the corresponding masses to high precision. 
Chargino, neutralino, slepton and sneutrino pair production cross section measurements would probe the loop corrections to gauge couplings in the supersymmetric sector. A $\mu^+\mu^-\to Z^0$ factory, utilizing the partial polarization of the muons, could allow significant improvements in $\sin^2\theta_{\rm w}$ precision and in $B$-mixing and CP-violating studies. 

The NMC will be particularly valuable for reconstructing supersymmetric particles of high mass from their complex cascade decay chains. Also, any $Z'$ resonances within the kinematic reach of the machine would give enormous event rates. The effects of virtual $Z'$ states would be detectable to high mass. If no Higgs bosons exist below $\sim$1~TeV, then the NMC would be the ideal machine for the study of strong $WW$ scattering at TeV energies.

Plus, there are numerous other new physics possibilities for muon facilities that are beyond the scope of the present report\cite{fermi}. In the following sections the physics opportunities above are discussed in greater detail. The work on physics at muon colliders reported in Sections~III, VI and IX is largely based on collaborations with M.S.~Berger, J.F.~Gunion and T.~Han.

\section{Front End Physics}

\subsection{Rare muon decays}

The planned muon flux ${\sim}10^{14}$ muons/sec for a muon collider dramatically eclipses the flux ${\sim}10^8$ muons/sec of present sources. With an intense source the rare muon processes $\mu\to e\gamma$ (current branching fraction $<0.49\times10^{-12}$), $\mu N\to eN$ conversion, and the muon electric dipole moment can be probed at very interesting levels. A generic prediction of supersymmetric grand unified theories is that these lepton flavor violating or CP violating processes should occur via loops at significant rates, e.g.\ BF$(\mu\to e\gamma)\sim 10^{-13}$\cite{bhs}. Lepton flavor violation can also occur via $Z'$ bosons, lepton quarks, and heavy neutrinos\cite{marciano}. 

\subsection{$\mu p$ collider}

The possibility of colliding 200~GeV muons with 1000~GeV protons at Fermilab is under study\cite{schellman}. This collider would reach a maximum $Q^2\sim8\times10^4$~GeV$^2$, which is $\sim$90~times the reach of the HERA $ep$ collider, and deliver a luminosity ${\sim}10^{33}\rm\,cm^{-2}\,s^{-1}$, which is $\sim$300 times the HERA luminosity. The $\mu  p $ collider would produce ${\sim}10^6$ neutral current deep inelastic scattering events per year at $Q^2>5000\rm~GeV^2$, which is over a factor $10^3$ higher than at HERA. This $\mu p$ collider would have a sensitivity to probe leptoquarks up to a mass $M_{LQ} \sim 800$~GeV and contact interactions to a scale $\Lambda\sim6$--9~TeV\cite{cheung}.

\subsection{Neutrino flux}

Muon decays are the way to make neutrino beams of well defined flavors\cite{schellman,fk}. A muon collider would yield a neutrino flux 1000 times that of the presently available neutrino flux. Then ${\sim}10^6$ $\nu N$ and $\bar\nu N$ events per year could be obtained to measure charm production ($\sim$6\% of the total cross section) and to measure $\sin^2\theta_{\rm w}$ (and infer the $W$-mass to an accuracy $\Delta M_W \simeq 30$--50~MeV in one year)\cite{schellman}.

\subsection{Neutrino oscillations}

A special purpose muon ring has been proposed by S.~Geer\cite{geer} to store ${\sim}10^{21}$ $\mu^+$ or $\mu^-$ per year and obtain ${\sim}10^{20}$ neutrinos per year from muon decays along $\sim$75\,m straight sections of the ring, which would be pointed towards a distant neutrino detector. One would have known neutrino fluxes from $\mu^-\to\nu_\mu\bar\nu_e e^-$ or from $\mu^+\to\bar\nu_\mu\nu_e e^+$ decays. Then, for example, from the decays of stored $\mu^-$, the following neutrino oscillation channels could be studied by detection of the charged leptons from the interactions of neutrinos in the detector:
\[ \begin{array}{cc}
\underline{\rm\ oscillation\ }& \underline{\rm\ detect\ }\\
\nu_\mu\to\nu_e& e^-\\
\nu_\mu\to\nu_\tau& \tau^-\\
\bar\nu_e\to\bar\nu_\mu& \mu^+\\
\bar\nu_e\to\bar\nu_\tau& \tau^+
\end{array}
\]
The detected $e^-$ or $\mu^+$ have the ``wrong sign" from the leptons produced by the interactions of the $\bar\nu_e$ and $\nu_\mu$ flux. The known neutrino fluxes from muon decays could be used for long-baseline oscillation experiments at any detector on Earth. The probabilities for vacuum oscillations between two neutrino flavors are given by
\begin{equation}
P(\nu_a\to\nu_b) = \sin^2 2\theta \, \sin^2(1.27\delta m^2 L/E)
\end{equation}
with $\delta m^2$ in eV$^2$ and $L/E$ in km/GeV.  In a very long baseline experiment from Fermilab to Gran Sasso laboratory ($L=9900$~km) with $\nu$-energies $E_\nu=10$ to 50~GeV ($L/E = 1000$--200~km/GeV),  neutrino charged current interaction rates ${\sim}10^3$/year would result. Such an experiment would have sensitivity to oscillations down to $\delta m^2\sim10^{-5}\rm\,eV^2$ for $\sin^22\theta=1$\cite{geer}.

\section{Higgs Particles}

The MSSM has five Higgs bosons: $h^0,\, H^0,\, A^0,\, H^\pm$. The mass of the lightest neutral Higgs boson is bounded from above by $m_h\lsim130$~GeV and accordingly is the ``jewel in the SUSY crown''. Global analyses of precision electroweak data now indicate a preference for a light SM Higgs boson. Davier-H\"ocker\cite{d-h} infer $m_h=129^{+103}_{\phantom0-62}$~GeV and Erler-Langacker\cite{e-l} obtain $m_h =122^{+134}_{\phantom0-77}$~GeV. Since in the decoupling limit\cite{decoupl} the couplings of the SM and SUSY Higgs bosons are approximately equal, these findings may be the ``smoking gun" for the SUSY Higgs boson.

The goals of a muon collider for the Higgs sector are to precisely determine the light Higgs mass, width, and branching fractions, to differentiate the $h_{\rm MSSM}$ from the $h_{\rm SM}$, and to find and study the heavy neutral Higgs boson $H^0$ and $A^0$.
The production of Higgs bosons in the $s$-channel with interesting rates is an unique feature of a muon collider\cite{ourPR,bbgh}. The resonance cross section is
\begin{equation}
\sigma_h(\sqrt s) = {4\pi \Gamma(h\to\mu\bar\mu) \, \Gamma(h\to X)\over
\left(\hat s - m_h^2\right)^2 + m_h^2 \left(\Gamma_{\rm tot}^h \right)^2}
\end{equation}
Gaussian beams with root mean square resolution  down to $R=0.003\%$ are realizable\cite{palmer}. The corresponding root mean square spread $\srts$ in c.m.\ energy is
\begin{equation}
\srts = (2{\rm~MeV}) \left( R\over 0.003\%\right) \left(\sqrt s\over 100\rm~GeV\right) \,.
\end{equation}
The effective $s$-channel Higgs cross section convolved with a Gaussian spread
\begin{equation}
\bar\sigma_h(\sqrt s) = {1\over \sqrt{2\pi}\,\srts} \; \int \sigma_h (\sqrt{\hat s}) \; \exp\left[ -\left( \sqrt{\hat s} - \sqrt s\right)^2 \over 2\sigma_{\sqrt s}^2 \right] d \sqrt{\hat s} 
\end{equation}
is illustrated in Fig.~1 for $m_h = 110$~GeV, $\Gamma_h = 2.5$~MeV, and resolutions $R=0.01\%$, 0.06\% and 0.1\%\cite{ourPR,bbgh}. A resolution $\srts \sim \Gamma_h$ is needed to be sensitive to the Higgs width. The light Higgs width is predicted to be\cite{bbgh}
\begin{equation}
\begin{array}{lll}
\Gamma \approx 2\mbox{ to 3 MeV}& \rm if& \tan\beta\sim1.8\\
\Gamma \approx 2\mbox{ to 800 MeV}& \rm if& \tan\beta\sim20
\end{array}
\end{equation}
for $80{\rm~GeV}\lsim m_h\lsim120$~GeV.

\begin{figure}[b]
\centering\leavevmode
\epsfxsize=3.4in\epsffile{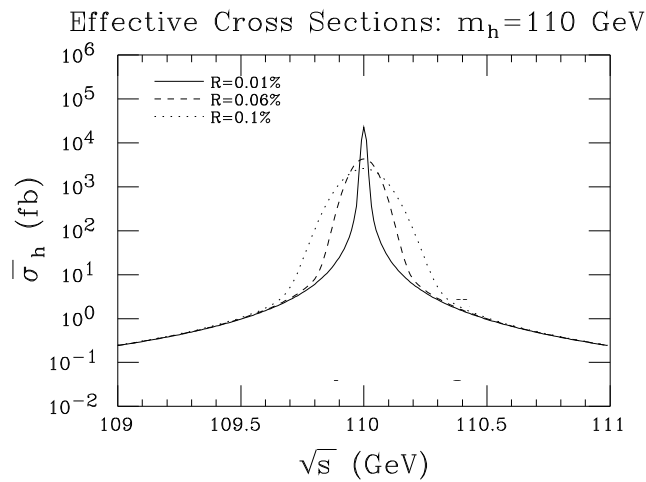}

\caption[]{Effective $s$-channel higgs cross section $\bar\sigma_h$ obtained by convoluting the Breit-Wigner resonance formula with a Gaussian distribution for resolution $R$. From Ref.~\cite{ourPR}.}
\end{figure}

At $\sqrt s = m_h$, the effective $s$-channel Higgs cross section is\cite{ourPR}
\begin{equation}
\bar\sigma_h \simeq {4\pi\over m_h^2} \; {{\rm BF}(h\to\mu\bar\mu) \,
{\rm BF}(h\to X) \over \left[ 1 + {8\over\pi} \left(\srts\over\Gamma_{\rm tot}^h\right)^2 \right]^{1/2}} \,.
\end{equation}
Note that $\bar\sigma_h\propto 1/\srts$ for $\srts>\Gamma_{\rm tot}^h$. At $\sqrt s = m_h \approx 110$~GeV, the $b\bar b$ rates are\cite{ourPR,bbgh}
\begin{eqnarray}
\rm signal &\approx& 10^4\rm\ events/fb\\
\rm background &\approx& 10^4\rm\ events/fb
\end{eqnarray}
assuming a $b$-tagging efficiency $\epsilon \sim 0.5$. The effective on-resonance cross sections for other $m_h$ values and other channels ($ZZ^*, WW^*$) are shown in Fig.~2 for the SM Higgs. The rates for the MSSM Higgs are nearly the same as the SM rates in the decoupling regime\cite{decoupl}, which is relevant at $\tan\beta\sim1.8$ in mSUGRA, corresponding to the infrared fixed point of the top quark Yukawa coupling\cite{bbo}.

\begin{figure}[h]
\centering\leavevmode
\epsfxsize=4in\epsffile{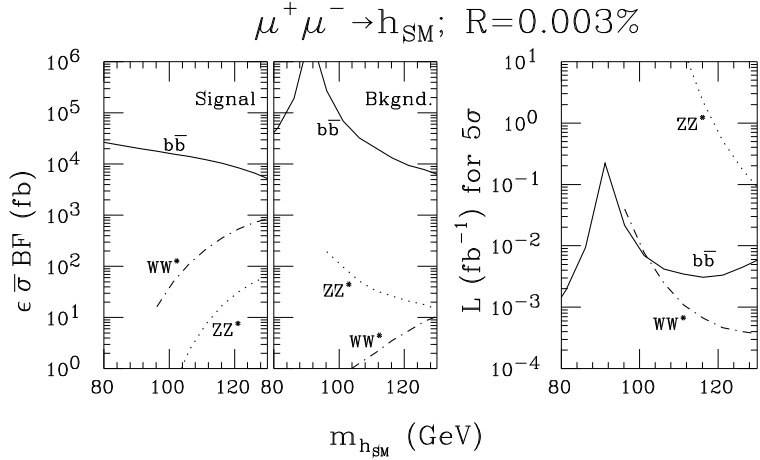}

\caption[]{The SM Higgs cross sections and backgrounds in $b\bar b,\ WW^*$ and $ZZ^*$. Also shown is the luminosity needed for a 5~standard deviation detection in $b\bar b$. From Ref.~\cite{ourPR}.}
\end{figure}

The important factors that make $s$-channel Higgs physics studies possible at a muon collider are energy resolutions $\srts$ of order a few MeV, little bremsstrahlung and no beamstrahlung smearing, and precise tuning of the beam energy to an accuracy $\Delta E\sim10^{-6}E$ through continuous spin-rotation measurements\cite{raja}. As a case study we consider $m_h \approx 110$~GeV. Prior Higgs discovery is assumed at the Tevatron (in $Wh, t\bar th$ production with $h\to b\bar b$ decay) or at the LHC (in $gg\to h$ production with $h\to \gamma\gamma, 4\ell$ decays with a mass measurement of $\Delta m_h \sim 100$~MeV for an integrated luminosity of $L=300\rm~fb^{-1}$) or possibly at a NLC (in $Z^*\to Zh, h\to b\bar b$ giving $\Delta m_h \sim 50$~MeV for $L=200\rm~fb^{-1}$). A muon collider ring design would be optimized to run at energy $\sqrt s= m_h$. For an initial Higgs mass uncertainty of $\Delta m_h\sim 100$~MeV, the maximum number of scan points required to locate the $s$-channel resonance peak at the muon collider is
\begin{equation}
n = 2\Delta m/\srts \approx 100
\end{equation}
for a resolution $\srts \approx 2$~MeV. The necessary luminosity per scan point ($L_{\rm s.p.}$) to observe or eliminate the $h$-resonance at a significance level $S/\sqrt B = 3$ is $L_{\rm s.p.} \sim 1.5\times10^{-3}\,\rm fb^{-1}$. (The scan luminosity requirements increase for $m_h$ closer to $M_Z$; at $m_h\sim M_Z$ the $L_{\rm s.p.}$ needed is a factor of 50 higher.) The total luminosity then needed to tune to a Higgs boson with $m_h = 110$~GeV is $L_{\rm tot} = 0.15\rm~fb^{-1}$. If the machine delivers  $1.5\times10^{31}\rm\, cm^{-2}\, s^{-1}$ (0.15~fb$^{-1}$/year), then one year of running  would suffice to complete the scan and measure the Higgs mass to an accuracy  $\Delta m \sim 1$~MeV. Figure~3 illustrates a simulation of such a scan. 

\begin{figure}[h]
\centering\leavevmode
\epsfxsize=3in\epsffile{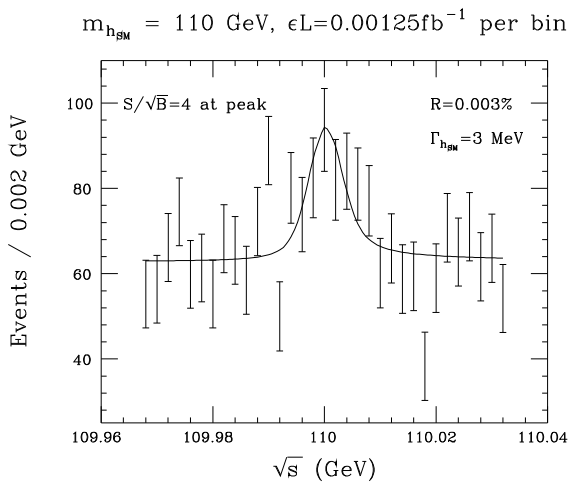}

\caption[]{Number of events and statistical errors in the $b\bar b$ final states as a function of $\sqrt s$ in the vicinity of $m_{h_{\rm SM}}=100$~GeV, assuming $R=0.03\%$. From Ref.~\cite{ourPR}.}
\end{figure}

Once the $h$-mass is determined to $\sim1$~MeV, a 3-point fine scan\cite{ourPR} can be made across the peak with higher luminosity, distributed with $L_1$ at the observed peak position in $\sqrt s$ and $2.5L_1$ at the wings ($\sqrt s = {\rm peak} \pm 2\srts$). Then with $L_{\rm tot}= 0.4\rm~fb^{-1}$ the following accuracies would be achievable: 16\% for $\Gamma_{\rm tot}^h$, 1\% for $\sigma\rm BF(b \bar b)$ and 5\% for $\sigma\rm BF(WW^*)$. The ratio $r = {\rm BF}(WW^*)/ {\rm BF} (b\bar b)$ is sensitive to $m_A$ for $m_A$ values below 500~GeV. For example, $r_{\rm MSSM}/r_{\rm SM} = 0.3, 0.5, 0.8$ for $m_A = 200, 250, 400$~GeV\cite{ourPR}. Thus, it may be possible to infer $m_A$ from $s$-channel measurements of $h$.

The study of the other neutral MSSM Higgs bosons at a muon collider via the $s$-channel is also of major interest. Finding the $H^0$ and $A^0$ may not be easy at other colliders. At the LHC the region $m_A>200$~GeV is deemed to be inaccessible for $3\lsim\tan\beta\lsim5$--10\cite{froid}. At an NLC the $e^+e^-\to H^0 A^0$ production process may be kinematically inaccessible if $H^0$ and $A^0$ are heavy. At a $\gamma\gamma$ collider, very high luminosity (${\sim}200\rm\ fb^{-1}$) would be needed for $\gamma\gamma\to H^0, A^0$ studies.

At a muon collider the resolution requirements for $s$-channel $H^0$ and $A^0$ studies are not as demanding as for the $h$, because the $H^0, A^0$ widths are broader; typically $\Gamma\sim30$~MeV for $m_A<2m_t$ and $\Gamma\sim3$~GeV for $m_A>2m_t$. Consequently $R\sim0.1\%$ ($\srts \sim 70$~MeV) is adequate for a scan. A luminosity per scan point $L_{\rm s.p.}\sim 0.1\rm~fb^{-1}$ probes the parameter space with $\tan\beta>2$.
 The $\sqrt s$-range over which the scan should be made depends on other information available to indicate the $A^0$ and $H^0$ mass ranges of interest.

In mSUGRA, $m_{A^0}\approx m_{H^0}\approx m_{H^\pm}$ at large $m_A$, with a very close degeneracy in these masses for large $\tan\beta$. In such a circumstance only an $s$-channel scan with good resolution may allow separation of the $A^0$ and $H^0$ states; see Fig.~4.

\begin{figure}[h]
\centering\leavevmode
\epsfxsize=3in\epsffile{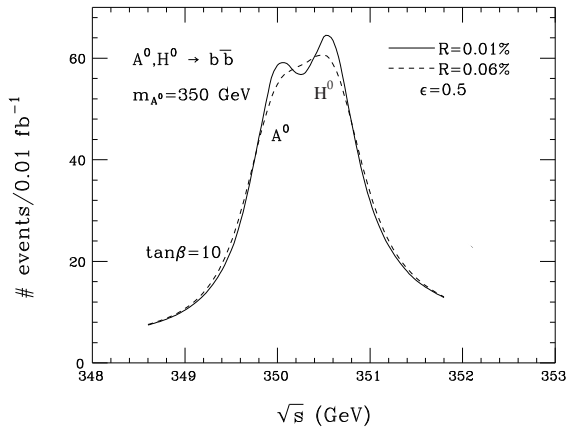}

\caption[]{Separation of $A^0$ and $H^0$ signals for $\tan\beta=10$. From Ref.~\cite{ourPR}. }
\end{figure}

\section{Technicolor Particles}

In modern top-assisted technicolor models\cite{techni}, the lighter neutral technipion resonances are expected to have masses in the range 100 to 500~GeV and widths of order 0.1 to 50~GeV\cite{bhat}. These resonances would be produced in the $s$-channel at a muon collider,
\begin{equation}
\mu^+\mu^-\to\pi^0_T, \, \rho^0_T,\, \omega^0_T 
\end{equation}
with high event rates.
The peak cross sections for these processes are estimated to be $\approx10^7$--$10^4$~fb\cite{bhat}. The dominant decay modes are\cite{bhat}
\begin{eqnarray}
\pi^0_T &\to& b\bar b, \, \tau\bar\tau,\, c\bar c,\, t\bar t\,,\\
\pi_T^{0\prime} &\to& gg,\, b\bar b,\, c\bar c,\, t\bar t,\, \tau^+\tau^-\,,\\
\rho^0_T &\to& \pi_T\pi_T,\, W\pi_T,\, WW \,,\\
\omega^0_T &\to& c\bar c,\, b\bar b,\, \tau\bar\tau,\, t\bar t,\, \gamma\pi^0_T,\, Z\pi^0_T \,.
\end{eqnarray}
Such resonances would be easy to find and study at a muon collider.

\section{$Z$-Factory}

A muon collider operating at the $Z$-boson resonance energy is an interesting option for measurement of polarization asymmetries, $B_s^0$--$\bar B_s^0$ mixing, and of CP violation in the $B$-meson system\cite{demarteau}. The muon collider advantages are the partial muon beam polarization, the separation of $b$ and $\bar b$ in $Z\to b\bar b$ events, and the long $B$-decay length for $B$-mesons produced at this $\sqrt s$. The left-right asymmetry $A_{LR}$ is the most accurate measure of $\sin^2\theta_{\rm w}$, since the uncertainty is statistics dominated. The present LEP and SLD polarization measurements show standard deviations of 2.4 in $A_{LR}^0$, 1.9 in $A_{FB}^{0,b}$ and 1.7 in $A_{FB}^{0,\tau}$\cite{e-l}. The CP angle $\beta$ (see Fig.~5) could be measured from $B^0\to K_s J/\psi$ decays. To achieve significant improvements over existing measurements and those at future $B$-facilities, more than $10^7$ $Z$-boson events would be needed, corresponding to a luminosity $>0.15\rm~fb^{-1}$, within the domain of muon collider expectations.

\begin{figure}[h]
\centering\leavevmode
\epsfxsize=1.75in\epsffile{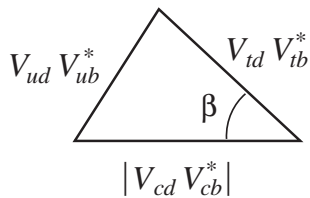}

\medskip
\caption{ Unitarity triangle for 3-generation mixing.}
\end{figure}

\section{Threshold Measurements at a\hfil\break Muon Collider}

With 10~fb$^{-1}$ integrated luminosity devoted to a measurement of a threshold cross-section, the following precisions on particle masses may be achievable\cite{bbgh2,bbh-new}:
\begin{equation}
\begin{array}{ll}
\mu^+\mu^-\to W^+W^-& \Delta M_W = 20\rm\ MeV\,,\\
\mu^+\mu^-\to t\bar t & \Delta m_t = 0.2\rm\ GeV\,,\\
\mu^+\mu^-\to Zh& \Delta m_h = 140{\rm\ MeV\ \ (if\ } m_h = 100\rm\ GeV)\,.
\end{array}
\end{equation}
Precision $M_W$ and $m_t$ measurements allow important tests of electroweak radiative corrections through the relation
\begin{equation}
M_W = M_Z \left[ 1 - {\pi\alpha\over \sqrt 2 \, G_\mu \, M_W^2 (1-\delta r)} \right]^{1/2} \,,
\end{equation}
where $\delta r$ represents loop corrections. In the SM, $\delta r$ depends on $m_t^2$ and $\log m_h$. The optimal precision for tests of this relation is $\Delta M_W \approx {1\over 140}\Delta m_t$, so the uncertainty on $M_W$ is the most critical. With $\Delta M_W=20$~MeV the SM Higgs mass could be inferred to an accuracy 
\begin{equation}
\Delta m_{h_{\rm SM}} = \pm 30{\rm\ GeV} \left(m_h\over 100\rm\ GeV\right)\,.
\end{equation}
Alternatively, once $m_h$ is known from direct measurements, SUSY loop contributions can be tested.

In top quark production at a muon collider above the threshold region, modest muon polarization would allow sensitive tests of anomalous top quark \mbox{couplings\cite{parke}}.

One of the important physics opportunities for the First Muon Collider is the production of the lighter chargino, $\tilde\chi_1^+$\cite{carprot}. Fine tuning arguments in mSUGRA suggest that it should be lighter than 200~GeV\cite{chen}. A search at the upgraded Tevatron for the process $q\bar q\to\tilde\chi_1^+\tilde\chi_2^0$ with $\tilde\chi_1^+\to \tilde\chi_1^0\ell^+\nu$ and $\tilde\chi_2^0\to\tilde\chi_1^0\ell^+\ell^-$ decays can potentially reach masses $m_{\tilde\chi_1^+}\simeq m_{\tilde\chi_2^0}\sim 170$~GeV with 2~fb$^{-1}$ luminosity and $\sim230$~GeV with 10~fb$^{-1}$\cite{teva2000}. The mass difference $M(\tilde\chi_2^0) - M(\tilde\chi_1^0)$ can be determined from the $\ell^+\ell^-$ mass distribution.

The two contributing diagrams in the chargino pair production process are shown in Fig.~6; the two amplitudes interfere destructively\cite{feng}. The $\tilde\chi_1^+$  and $\tilde\nu_\mu$ masses can be inferred from the shape of the cross section in the threshold region\cite{bbh-new}. The chargino decay is $\tilde\chi_1^+\to f\bar f' \tilde\chi_1^0$. Selective cuts suppress the background from $W^+W^-$ production and leave $\sim5\%$ signal efficiency for 4\,jets${}+\E$ events. Measurements at two energies in the threshold region with total luminosity $L=50\rm~fb^{-1}$ and resolution $R=0.1\%$ can give the accuracies listed in Table~1 on the chargino mass for the specified values of $m_{\tilde\chi_1^+}$ and $m_{\tilde\nu_\mu}$. 

\begin{figure}[h]
\centering\leavevmode
\epsfxsize=3.5in\epsffile{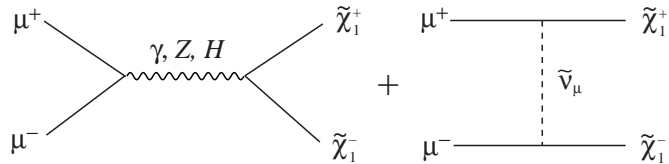}

\caption{Diagrams for production of the lighter chargino.}
\end{figure}

\begin{table}[b]
\caption[]{Achievable uncertainties with 50~fb$^{-1}$ luminosity on the mass of the lighter chargino for representative $m_{\tilde\chi_1^+}$ and $m_{\tilde\nu_\mu}$ masses. From Ref.~\cite{bbh-new}. }
\begin{tabular}{c}
\hfill\epsfxsize=3in\epsffile{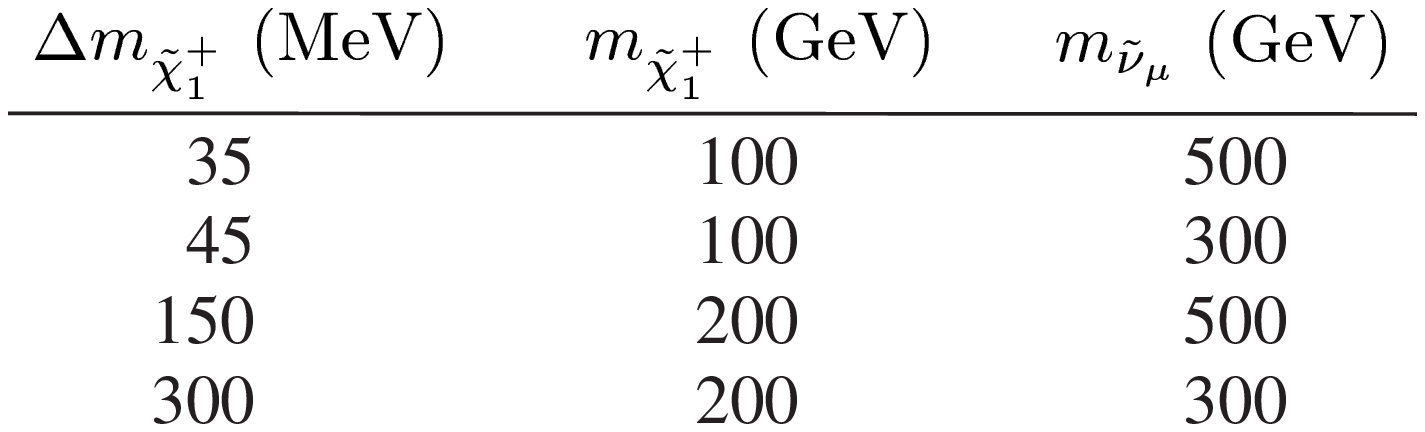}\hfil
\end{tabular}
\end{table}

\section{Supersymmetric Radiative Corrections}

In unbroken supersymmetry, the SUSY gaugino couplings $h_i$ to $\tilde ff$ are equal to the SM gauge couplings $g_i$. In broken SUSY a difference in $h_i$ and $g_i$ couplings is induced at the loop level due to different mass scales for squarks and sleptons\cite{chank,nojiri,cheng,diaz}. The differences in the U(1) and SU(2) couplings are\cite{cheng}
\begin{eqnarray}
{h_1-g_1\over g_1} &\simeq& 1.8\% \log_{10} \left(M_{\tilde Q}\over m_{\tilde \ell} \right) \,,\\
{h_2-g_2\over g_2} &\simeq& 0.7\% \log_{10} \left(M_{\tilde Q}\over m_{\tilde \ell}\right) \,.
\end{eqnarray}
One-loop amplitudes for SUSY processes are obtained from the tree-level amplitudes by substitution of the modified couplings. The cross-sections of SUSY processes with $t$-channel exchanges can be enhanced up to $9\%\log_{10} \left( M_{\tilde Q}/m_{\tilde\ell}\right)$\cite{nojiri}. Consequently, precision cross-section measurements can be sensitive to squarks of mass $M_{\tilde Q}>1$~TeV. If the first two generations have masses in the 1 to 40~TeV range allowed by naturalness, then precision measurements could provide a way to infer squark masses beyond the kinematic reach of colliders.

Some $t$-channel exchange processes of interest in this regard at muon colliders are shown in Fig.~7. The technique relies on knowledge of the exchanged particle mass, which must be determined from its production processes. The muon collider advantage in the study of supersymmetric radiative corrections is the accuracy with which mass measurements can be made near thresholds. 

\begin{figure}[h]
\centering\leavevmode
\epsfxsize=4in\epsffile{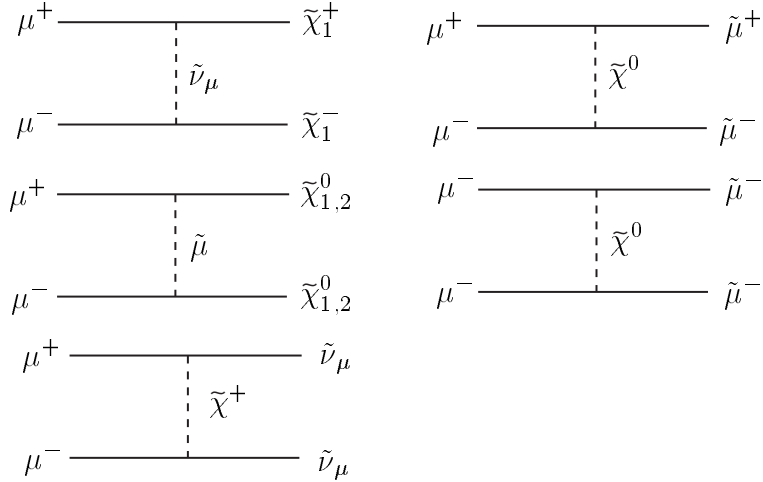}

\caption{$t$-channel exchange diagrams for processes that can be enhanced by SUSY radiative corrections.}
\end{figure}

\section{Heavy Particles of Supersymmetry}

The requirements of gauge coupling unification can be used to predict the mean SUSY mass scale, given the value of the strong coupling at the $Z$-mass scale. Figure~8 shows the SUSY GUT predictions versus $\alpha_s(M_Z)$. For the value $\alpha_s(M_Z) = 0.1214\pm0.0031$ from a new global fit to precision electroweak data\cite{e-l}, a mean SUSY mass of order 1~TeV is expected. Thus some SUSY particles will likely have masses at the TeV scale. 

\begin{figure}[h]
\centering\leavevmode
\epsfxsize=4.7in\epsffile{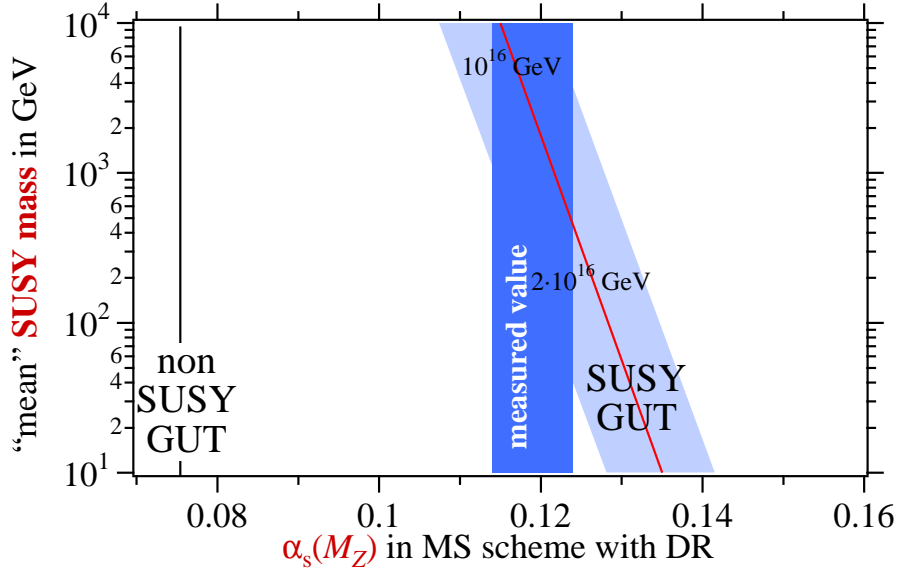}

\bigskip
\caption[]{$\alpha_s$ prediction in supersymmetric GUT with minimal particle content. From Ref.~\cite{barbi}.}
\end{figure}

At the LHC, mainly squarks and gluinos are produced; these decay to lighter SUSY particles. The LHC is a great SUSY machine, but some sparticle measurements will be very difficult or impossible there\cite{hinch,paige}, namely: (i)~the determination of the LSP mass (LHC measurements give SUSY mass differences); (ii)~study of sleptons of mass $\gsim200$~GeV because Drell-Yan production becomes too small at these masses; (iii)~study of heavy gauginos $\tilde\chi_2^\pm$ and $\tilde\chi_{3,4}^0$, which are mainly Higgsino and have small direct production rates and small branching fractions to channels usable for detection; (iv)~study of heavy Higgs bosons $H^\pm,\ H^0,\ A^0$ that have small cross sections and decays to $t\bar t$ that are likely dominant (their detection is deemed impossible if SUSY decays dominate).

With supersymmetry there will be many scalar particles. Pair production of scalar particles at a lepton collider is $P$-wave suppressed. Consequently, energies well above threshold are needed for sufficient production rates; see Fig.~9. A 3 to 4 TeV muon collider with high luminosity ($L\sim10^2$ to $10^3\rm~fb^{-1}/year$) could provide sufficient event rates to reconstruct heavy sparticles from their complex cascade decay chains\cite{paige,lykken}. 

\begin{figure}[t]
\centering\leavevmode
\epsfxsize=5in\epsffile{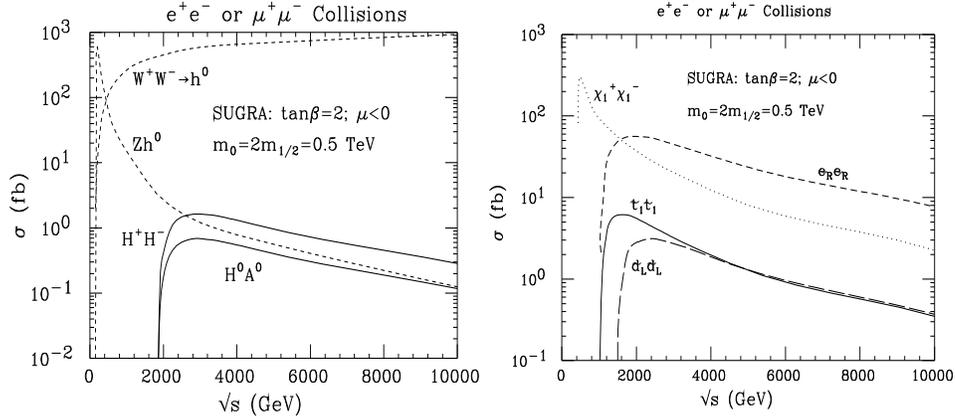}

\caption[]{Cross sections for pair production of Higgs bosons and scalar particles at a high energy muon collider. From Ref.~\cite{sanfran95}.}
\end{figure}

Extra $Z$ bosons and a low-energy supersymmetry are natural in string models\cite{lang-cvet}. The $s$-channel production of a $Z'$ boson at the resonance energy would give enormous event rates at the NMC.  Moreover, the $s$-channel contributions of $Z'$ bosons with mass far above the kinematic reach of the collider could be revealed as contact interactions\cite{god}.

\section{Strong Scattering of Weak Bosons}

The scattering of weak bosons can be studied at a high energy muon collider through the process in Fig.~10. The amplitude for the scattering of longitudinally polarized $W$-bosons behaves like\cite{bagger}
\begin{equation}
A(W_LW_L\to W_LW_L) \sim m_H^2/v^2
\end{equation}
if there is a light Higgs boson, and
\begin{equation}
A(W_LW_L\to W_LW_L) \sim s_{WW}^{\vphantom y} /v^2
\end{equation}
if no light Higgs boson exists; here $s_{WW}^{\vphantom y}$ is the square of the $WW$ c.m.\ energy and $v=246$~GeV. In the latter scenario, partial wave unitarity of $W_LW_L\to W_LW_L$ requires that strong scattering of weak bosons occurs at the 1 to 2~TeV energy scale. Thus subprocess energies $\sqrt{\smash{s_{WW}^{\vphantom y}}}\gsim 1.5$~TeV are needed to probe strong $WW$ scattering effects. 

\begin{figure}[t]
\centering\leavevmode
\epsfxsize2.5in\epsffile{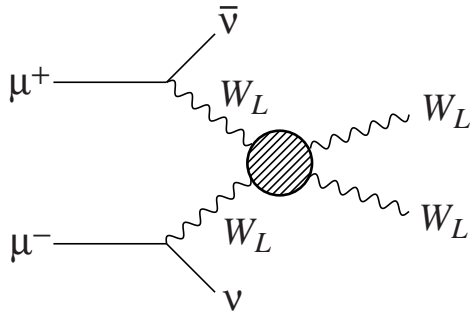}

\caption{Symbolic diagram for strong $WW$ scattering.}
\end{figure}

The nature of the dynamics in the $WW$ sector is unknown. Models for this
scattering assume heavy resonant particles (isospin scalar and vector) or a non-resonant amplitude which extrapolates the low-energy theorem behavior $A\sim s_{WW}^{\vphantom y}/v^2$. In all models, impressive signals of strong $WW$ scattering are obtained at the NMC, with cross sections typically of order 50~fb$^{-1}$\cite{W-mass}.

\section{Conclusions}

The Front End of a muon collider offers dramatic improvements in sensitivity for flavor-violating transitions (e.g.\ $\mu\to e\gamma$), high $Q^2$ phenomena in deep inelastic muon-proton and neutrino-proton interactions, and neutrino oscillation studies in long-baseline experiments.

The First Muon Collider offers unique probes of supersymmetry (particularly $s$-channel Higgs boson resonances), high precision threshold measurements of $W,\ t$ and SUSY particle masses, tests of SUSY radiative corrections that indirectly probe the existence of high mass squarks, and a possible $Z^0$ factory for improved precision in polarization measurements and for $B$-physics studies of CP violation and mixing.

The Next Muon Collider guarantees access to heavy SUSY scalar particles and $Z'$ states or to strong $WW$ scattering if there are no Higgs bosons and no supersymmetry.

The bottom line is that muon colliders are robust options for probing new physics that may not be accessible at other colliders.

\section*{Acknowledgments}
I would like to thank M.S.~Berger, G.~Burdman, J.F.~Gunion, T.~Han and C.~Kao for helpful advice in the preparation of this report. This research was supported in part by the U.S.~Department of Energy under Grant No.~DE-FG02-95ER40896 and in part by the University of Wisconsin Research Committee with funds granted by the Wisconsin Alumni Research Foundation.

\end{document}